\documentclass[aps,prl,reprint,superscriptaddress,longbibliography]{revtex4-1}

\usepackage{graphicx,amssymb,amsmath,xcolor}
\usepackage[normalem]{ulem}
\usepackage{color}

\begin{document}

%\preprint{}

\title{Isotropy of cosmic rays beyond $10^{20}$~eV favors their heavy mass composition}


%\author{}
%\thanks{}
%\affiliation{}


%\date{\today}

\author{R.U. Abbasi}
\affiliation{Department of Physics, Loyola University Chicago, Chicago, Illinois 60660, USA}

\author{Y. Abe}
\affiliation{Academic Assembly School of Science and Technology Institute of Engineering, Shinshu University, Nagano, Nagano 380-8554, Japan}

\author{T. Abu-Zayyad}
\affiliation{Department of Physics, Loyola University Chicago, Chicago, Illinois 60660, USA}
\affiliation{High Energy Astrophysics Institute and Department of Physics and Astronomy, University of Utah, Salt Lake City, Utah 84112-0830, USA}

\author{M. Allen}
\affiliation{High Energy Astrophysics Institute and Department of Physics and Astronomy, University of Utah, Salt Lake City, Utah 84112-0830, USA}

\author{Y. Arai}
\affiliation{Graduate School of Science, Osaka Metropolitan University, Sugimoto, Sumiyoshi, Osaka 558-8585, Japan}

\author{R. Arimura}
\affiliation{Graduate School of Science, Osaka Metropolitan University, Sugimoto, Sumiyoshi, Osaka 558-8585, Japan}

\author{E. Barcikowski}
\affiliation{High Energy Astrophysics Institute and Department of Physics and Astronomy, University of Utah, Salt Lake City, Utah 84112-0830, USA}

\author{J.W. Belz}
\affiliation{High Energy Astrophysics Institute and Department of Physics and Astronomy, University of Utah, Salt Lake City, Utah 84112-0830, USA}

\author{D.R. Bergman}
\affiliation{High Energy Astrophysics Institute and Department of Physics and Astronomy, University of Utah, Salt Lake City, Utah 84112-0830, USA}

\author{S.A. Blake}
\affiliation{High Energy Astrophysics Institute and Department of Physics and Astronomy, University of Utah, Salt Lake City, Utah 84112-0830, USA}

% ...

\collaboration{The Telescope Array Collaboration}
\noaffiliation

\begin{abstract}
 We report an estimation of the injected mass composition of ultra-high energy cosmic rays (UHECRs)
 at energies higher than $10$~EeV. The composition is inferred from an energy-dependent sky
 distribution of UHECR events observed by the Telescope Array surface detector by comparing it
 to the Large Scale Structure of the local Universe. In the case of negligible extra-galactic
 magnetic fields the results are consistent with a relatively heavy injected composition at $E \sim 10$~EeV
 that becomes lighter up to $E \sim 100$~EeV, while the composition at $E > 100$~EeV is very heavy.
 The latter is true even in the presence of highest experimentally allowed extra-galactic magnetic fields,
 while the composition at lower energies can be light if a strong EGMF is present.
 The effect of the uncertainty in the galactic magnetic field on these results is subdominant.
\end{abstract}

%\keywords{}

\maketitle

%\section{}
%\subsection{}
%\subsubsection{}

% General introduction
Ultra-high energy cosmic rays (UHECR) are charged particles, likely protons
and nuclei, with energies greater than $1$~EeV ($10^{18}$~eV) that are
reaching Earth's from space.  The flux of particles at these energies is tiny, of order
$1~{\rm km^{-2} sr^{-1} yr^{-1}}$, so they can be detected only indirectly via
extensive air showers (EAS) of secondary particles they initiate in the Earth
atmosphere. Despite several decades of study the origin of UHECR and the nature of their primary particles remain unknown. The UHECR energy spectrum
was measured with a good precision~\cite{PierreAuger:2010gfm, AbuZayyad:2012ru}; its general shape is consistent between the two modern experiments Pierre Auger
(Auger)~\cite{PierreAuger:2015eyc} and Telescope Array
(TA)~\cite{TelescopeArray:2012uws} 
and with theoretical models~\cite{Greisen:1966jv, Zatsepin:1966jv, Aloisio:2009sj},
except for a minor discrepancy~\cite{TelescopeArray:2021zox} at highest energies.  The spectrum
measurements alone, however, have a limited potential to discriminate between
various models of UHECR origin.
The mass composition measurements have generally better discriminating
power. But opposite to the
spectrum, the mass composition measurements of
Auger~\cite{PierreAuger:2014gko, PierreAuger:2017tlx} and TA~\cite{TelescopeArray:2018xyi, TelescopeArray:2018bep} are more affected by various systematic effects and not covering the highest energy part of the UHECR spectrum.
At the same time the UHECR arrival directions are measured with a sufficient
precision of order $1^\circ$.  Unfortunately, this does not allow one to
directly identify the sources since the deflections of UHECR are highly uncertain
because of both unknown event-by-event primary particle
charges, and because of large
uncertainties in the galactic and extragalactic magnetic fields.
Several approaches have been proposed in the literature to decipher the origin of UHECR using complex anisotropy observables~\cite{Kalashev:2019skq, Urban:2020szk, Bister:2020rfv, Abbasi:2020fxl}.

%General idea and proposal
In this letter we use a novel method to
infer the {\it injected} UHECR mass composition from the arrival directions of
the TA events.
The method was proposed and described in detail in Ref.~\cite{Kuznetsov:2020hso}.
It takes advantage of the accurate measurements of UHECR arrival directions and energy,
while circumventing the uncertainties arising from cosmic magnetic fields.
The method is based on the observation that the magnitude of UHECR deflections
is determined predominantly by particle charges that may range from 1 for
protons to 26 for iron, while other factors are expected to give an order of magnitude smaller
effect. Comparing the energy-dependent UHECR distribution over the sky
calculated with various injected mass compositions with the observed
distribution one may identify the models that are compatible or incompatible with 
the data. At this stage, the parameters of the UHECR models other than the mass composition
are fixed by some conservative assumptions. One may then vary these parameters
to check if the conclusions about the mass composition are robust with respect
to this variation.
Somewhat similar approaches to UHECR mass composition estimation from their anisotropy
have been proposed in Refs.~\cite{Anjos:2018mgr, Tanidis:2022jox}.

%Experiment and data
The Telescope Array~\cite{TelescopeArray:2012uws} is the largest cosmic-ray
experiment in the Northern Hemisphere.  It is located at $39.3^\circ$~N,
$112.9^\circ$~W in Utah, USA. The observatory includes a surface detector
array (SD) and $38$ fluorescence telescopes grouped in three stations. The
SD consists of $507$ plastic scintillator stations of 3~m$^2$ each, 
which are placed in a square grid with the $1.2$~km
spacing, covering in total the area of $\sim 700 \; {\rm km}^2$. The TA SD can 
detect EAS produced by cosmic ray particles of $\sim$EeV and
higher energies. The TA SD has been in operation since May 2008. In this analysis we use the
data collected by the TA SD during $14$ years of operation from May 11, 2008 to
May 10, 2022. We use the quality cuts described in
Ref.~\cite{TelescopeArray:2014ahm}, and select events with zenith angle $\theta
<55^\circ$ and energy $E > 10$~EeV.
We also use the data of the National Lightning Detection Network~\cite{NLDN} to
filter out the events possibly caused by lightnings as described in
Ref.~\cite{TelescopeArray:2018rbt}. The resulting data set contains 5978
events, including the event with the highest energy of 244~EeV~\cite{TelescopeArray:2023sbd} and $18$ other events with $E > 100$~EeV.

%Reconstruction
Each event that activates the SD trigger is recorded,  and the kinematic
parameters of its primary particle are reconstructed.  The arrival direction
is determined from the relative difference in arrival times of the shower front
at each surface detector, which is measured with the precision of $20$~ns.
The energy of the primary particle is estimated using the EAS particle density
$S_{800}$ measured at a distance of $800$~m from the shower axis.  The measured value of
$S_{800}$ is converted to the reconstructed SD energy taking into account the 
zenith angle dependence by means of a Monte-Carlo simulation that uses the CORSIKA software
package~\cite{Heck:1998vt}. Finally, thus reconstructed SD energy is calibrated to the
calorimetric energy measured by the fluorescence detectors; this amounts to a
rescaling by the factor of $1/1.27$~\cite{AbuZayyad:2012ru}. The resolution of
the SD at $E > 10$~EeV is $1.4^{\circ}$ in arrival direction and $18\%$ in
the logarithm of primary energy~\cite{AbuZayyad:2012ru, 2014arXiv1403.0644T}. The
systematic uncertainty in the energy determination is estimated at 
$21\%$~\cite{TheTelescopeArray:2015mgw}.

The implementation of our method is organized in three steps.  First, we
generate a large mock set of realistic UHECR events  
for each injected composition model considered. Second, we define
the test-statistics (TS) that quantifies the overall magnitude of 
deflections of a given event set with respect to the Large Scale Structure of
the Universe (LSS) and that is robust to the uncertainties of the
magnetic fields.  Finally, we calculate this TS for each mock event set as well as for the 
real data, and quantify the compatibility of each composition
model with the data. The effect of the uncertainties in magnetic fields and
injection spectra is estimated by varying their parameters for each 
composition model.

%Details of the mock sets simulations
We now describe these steps in more detail, starting with a brief description
of the key properties of the UHECR mock event sets; more thorough
discussion is given in a companion paper~\cite{TelescopeArray:2024buq}. We assume that UHECR
sources trace the matter distribution in the local Universe. Statistically,
this can be achieved by assuming equal intrinsic UHECR flux for each galaxy in
a complete volume-limited sample. In practice we use the flux-limited galaxy sample
with a high degree of completeness, derived from the 2MRS galaxy catalog
\cite{Huchra:2011ii} by cutting out galaxies with ${\rm mag} >
12.5$ and with distances below $5$~Mpc and beyond $250$~Mpc. 
We assign a progressively larger flux to more distant galaxies to
compensate for the observational selection inherent in a flux-limited
sample. The sources beyond $250$~Mpc are assumed to be distributed uniformly
with the same mean density as those within this distance. 
Their contribution is added as a properly normalized fraction of isotropic
events. The exact procedure is described in Ref.~\cite{Koers:2009pd}.  This source
model 
covers all the source scenarios with sufficiently numerous sources (source
number density $\rho \gg 10^{-5}$~Mpc$^{-3}$).
The source densities of order $10^{-5}$~Mpc$^{-3}$ are not excluded
experimentally~\cite{PierreAuger:2013waq} (see, however, recent studies~\cite{Kuznetsov:2023jfw, Bister:2023icg}). In this case the 
sensitivity of our method to the mass composition decreases; 
we discuss this issue in a companion paper~\cite{TelescopeArray:2024buq}.

We fix the injection spectrum for each nucleus by deriving it from the
separate fit to the TA and Auger observed spectra~\cite{diMatteo:2017dtg, TelescopeArray:2024buq}. As
a result the following spectra are taken for the UHECR flux simulation: power law
with the slope $-2.55$, $-2.20$, $-2.10$ and without the cut-off for {\it protons},
{\it helium} and {\it oxygen}, respectively; power law with the slope $-1.50$ and
with a sharp cut-off at $280$~EeV for {\it silicon}; power law with the slope
$-1.95$ and with a sharp cut-off at $560$~EeV for {\it iron}. 
The secondary particles produced upon propagation of injected primary nuclei through the
interstellar medium are taken into account for helium and oxygen nuclei and
reasonably neglected for other primaries; the details are given in
Ref~\cite{diMatteo:2017dtg}. We also consider separately a best-fit injected composition model from the Auger work~\cite{Aab:2016zth}, where we take into account all the secondaries and model the deflection of the full flux according to its average charge.

The deflections in magnetic fields are treated with the account of primary particle 
charge $Z$ and its energy $E$.  The deflections in the extra-galactic
magnetic field (EGMF) are simulated as a direction-independent smearing of the
sources with the von Mises-Fischer distribution. For our basic model its magnitude
is set to zero, which corresponds to either $B_{\rm EGMF} \ll 1$~nG for the
correlation length $\lambda \sim 1$~Mpc or $B_{\rm EGMF} \ll 0.1$~nG for $\lambda$
of a cosmological scale. We discuss the possible effect of non-zero EGMF
among other uncertainties. The deflections in the regular galactic magnetic field
(GMF) are simulated using the backtracking technique with the GMF model of
Ref.~\cite{Pshirkov:2011um}.  The deflections in the random
GMF are simulated as a galactic-latitude-dependent smearing according to the
data-driven relation of Ref.~\cite{Pshirkov:2013wka}. Finally, the event
distribution is modulated by the geometrical exposure of the TA. The energies of the events in the mock sets are generated according to the observed TA spectrum with the account of the TA energy resolution. In companion paper~\cite{TelescopeArray:2024buq} we estimate the
impact of uncertainties in the energy scale and in the parameters of the injection spectra and magnetic fields on the inferred mass composition.

%Details of the test-statistics
We define the test-statistics (TS) using the expected UHECR flux maps built by a
similar procedure as used for the mock sets generation, but with smaller
number of free parameters.  Namely, we use the same 2MRS-based source catalog,
assume flux attenuation as protons with $\sim E^{-2.55}$ injection spectrum
without cutoff and a uniform smearing of sources. The magnitude of this smearing $\theta_{100}$ 
defined at 100~EeV is the only free parameter on which the TS depends.
For each given value of $\theta_{100}$ we build a set of maps $\Phi_k(\theta_{100},{\bf n})$
where ${\bf n}$ is the direction in the sky, $k$ denotes the energy bin
and the smearing of each map scales properly, 
as $100\; {\rm EeV}/E_k$.
Then the test statistics ${\rm TS}(\theta_{100})$ for a given event set 
with directions ${\bf n}_i$ is defined as follows:
\begin{equation}
{\rm TS}(\theta_{100}) = -2 \sum_k \left( 
\sum_i \ln  \frac{\Phi_k(\theta_{100}, {\bf n}_i)} 
{\Phi_{\rm iso} ({\bf n}_i)}
\right), 
\label{eq:TS}
\end{equation}
where the sum run over the events $i$ and energy bins $k$, and we have
included a standard overall normalization factor $-2$. The normalization factor
$\Phi_{\rm iso}({\bf n}_i) = \Phi(\infty,{\bf n}_i)$ corresponding to 
an isotropic distribution is added for convenience. More technical details on the TS construction are given in the companion paper~\cite{TelescopeArray:2024buq}.
In the limit of a large number of events, this test statistics is distributed
around its minimum according to $\chi^2$-distribution with one degree of
freedom.  The position of the TS minimum $\theta_{100}^{\rm min}$ for each
event set is interpreted as the energy-rescaled mean event deflection with
respect to the LSS.  Thus, for a mock set of a given composition
model and a very large number of events, the TS should have a deep and narrow
minimum, with the value of $\theta_{100}^{\rm min}$ being characteristic of this
composition model.  These values could then be confronted with the ${\rm TS}(\theta_{100})$ evaluated for the data.

\begin{figure}%[ht]
 \includegraphics[width=0.99\columnwidth]{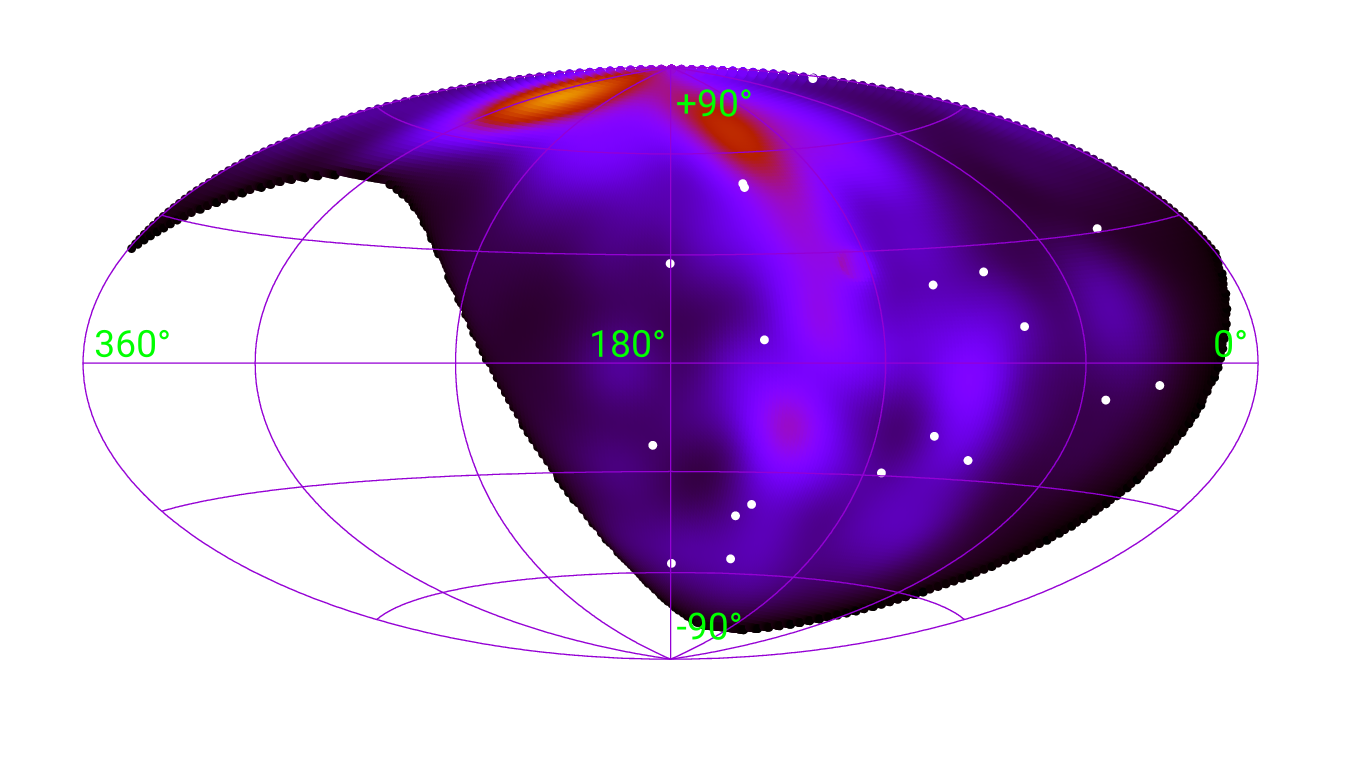}
 \includegraphics[width=0.99\columnwidth]{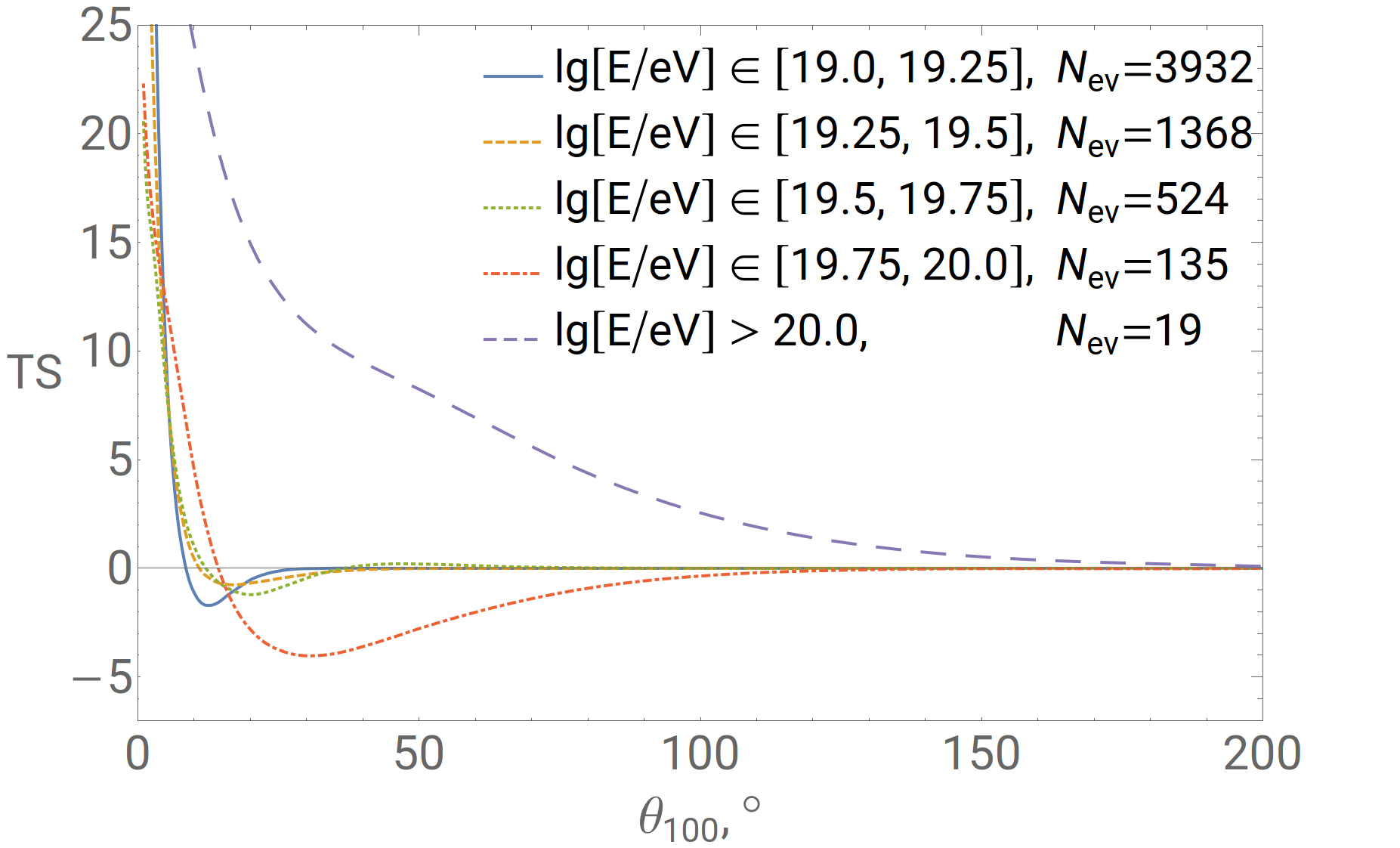}
\caption{
\label{fig:TS-data}
{\it Top panel:} Example of the map $\Phi_k$ ($E > 100$~EeV, $\theta_{100} = 10^\circ$) used for test-statistics computation, overlaid with the distribution of the TA SD events with $E > 100$~EeV (two of them are forming a doublet). The color indicates the expected distribution of the cosmic ray flux. Galactic coordinates.
{\it Bottom panel:}
The distribution of test-statistics over $\theta_{100}$
evaluated for experimental data in five energy bins. The number of events in each bin is shown, in the legend.}
\end{figure}
To estimate the mass composition we divide the energy range into 5 bins
starting from $10$~EeV with a quarter-decade width and with the last bin being
an open interval $E > 100$~EeV. The dependence of ${\rm TS}(\theta_{100})$ on $\theta_{100}$
for the data in each bin is shown in Fig.~\ref{fig:TS-data}. The
curves for all but the penultimate bin (red curve) are consistent, at the $2\sigma$ level, with isotropy which corresponds to $\theta_{100} = 200^\circ$ in our notations ---
the value that is beyond the size of the TA field of view. In the bin $19.75 <
\log_{10}[E/eV] < 20.0$ the TS has a distinct minimum at $\theta_{100}^{\rm
  min} = 30.8 ^\circ$ that deviates from isotropy with the significance of more than $2 \sigma$.

\begin{figure}
 \includegraphics[width=0.99\columnwidth]{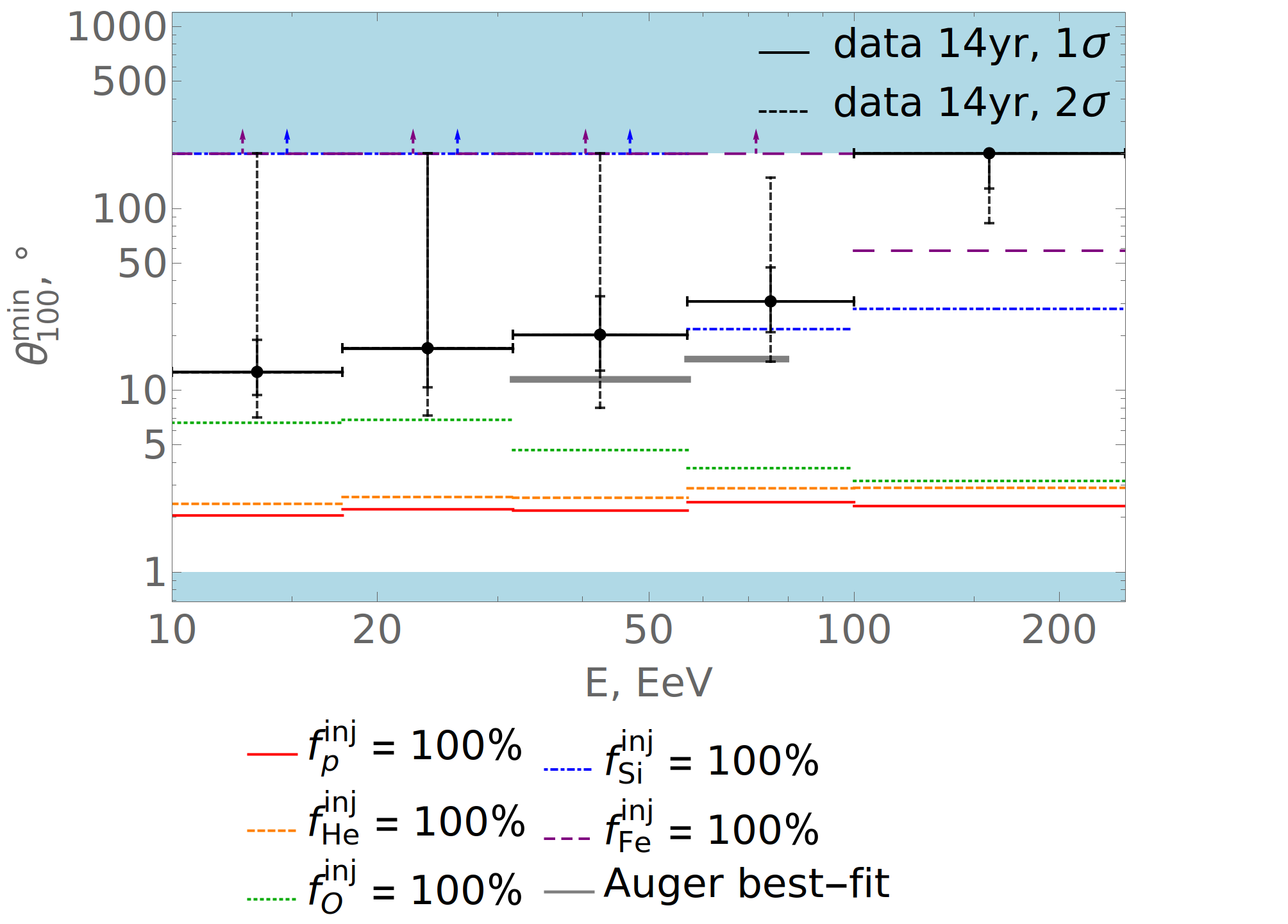}
\caption{
\label{fig:rails}
The distribution of the test-statistics minima, $\theta_{100}^{\rm min}$, for the data compared to several injected composition models.
Regular GMF model of Ref.~\cite{Pshirkov:2011um} is used, and deflections in EGMF
are neglected. Note that several composition models yield the same value of 
$\theta_{100}^{\rm min} = 200^\circ$, i.e. they are indistinguishable in our method.
The corresponding lines which merge together on the plot are indicated by arrows. Pure nuclei composition models and Auger best-fit composition model of Ref.~\cite{Aab:2016zth} (see text).}
\end{figure}

In Fig.~\ref{fig:rails} we present a bin-wise comparison of the data with
various composition models. The data points are in correspondence with the
${\rm TS}(\theta_{100})$ curves shown in Fig.~\ref{fig:TS-data}: the central
points show values of  $\theta_{100}^{\rm min}$ in each bin, while the
error bars represent $1\sigma$- and $2\sigma$-deviations 
from the minimum as calculated from the corresponding curve. 
It should be stressed that, by definition, 
the data points show typical deflections of
cosmic rays in the corresponding bin {\em rescaled to 
$E=100$~EeV}. While the energy dependence of deflections is taken into
account in this way, the other factors such as the difference in attenuation
at different energies (and, therefore, relative contribution of close and
distant sources) are not. Hence the variations of $\theta_{100}^{\rm min}$
from bin to bin. 
Regardless these variations, it is manifest in 
Fig.~\ref{fig:rails} that the small values of $\theta_{100}$ are not
compatible with the data at all energies, which is evident already in
Fig.~\ref{fig:TS-data} from the steep rise of the curves at small
$\theta_{100}$. 

\begin{figure*}
 \includegraphics[width=0.99\columnwidth]{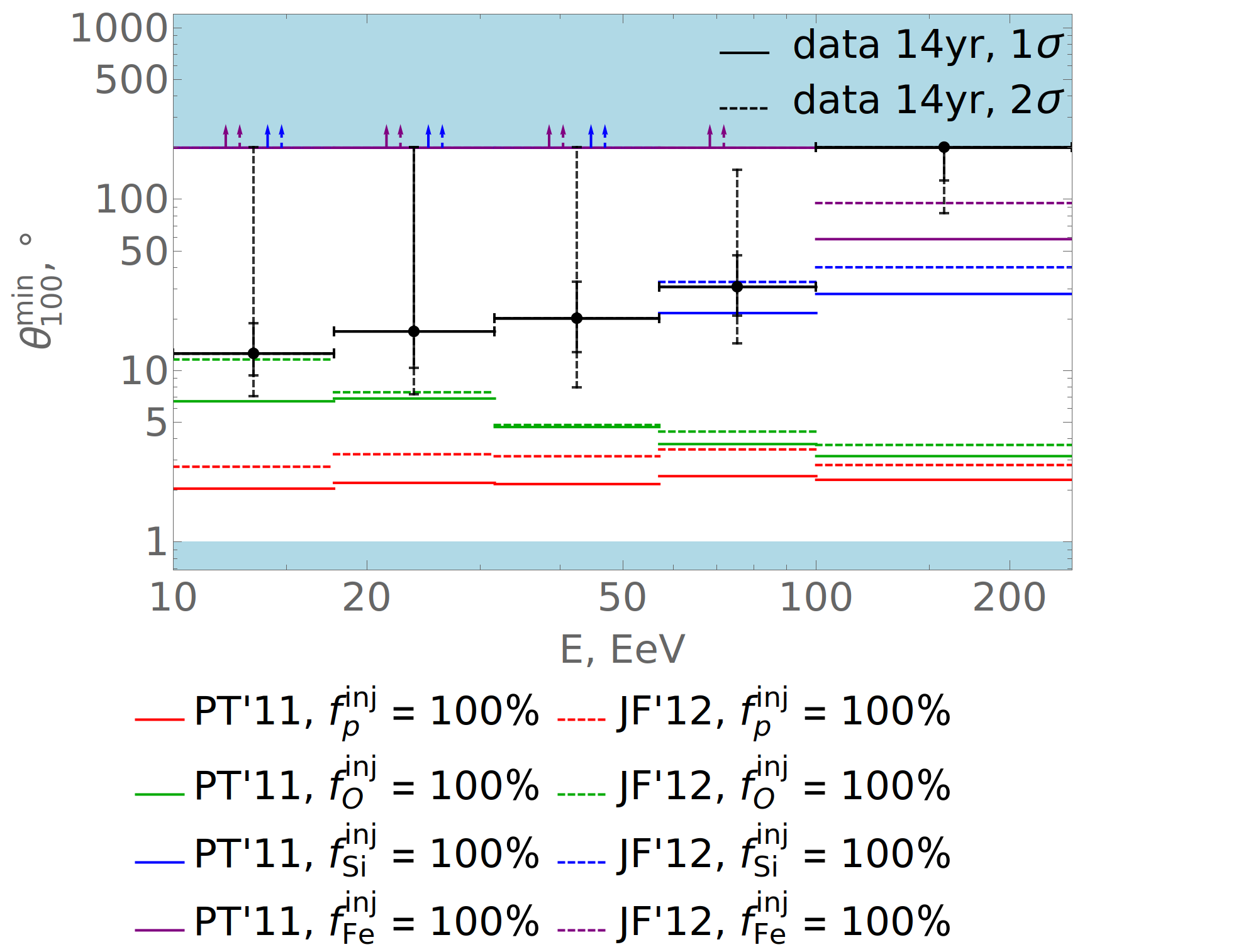}
 \includegraphics[width=0.99\columnwidth]{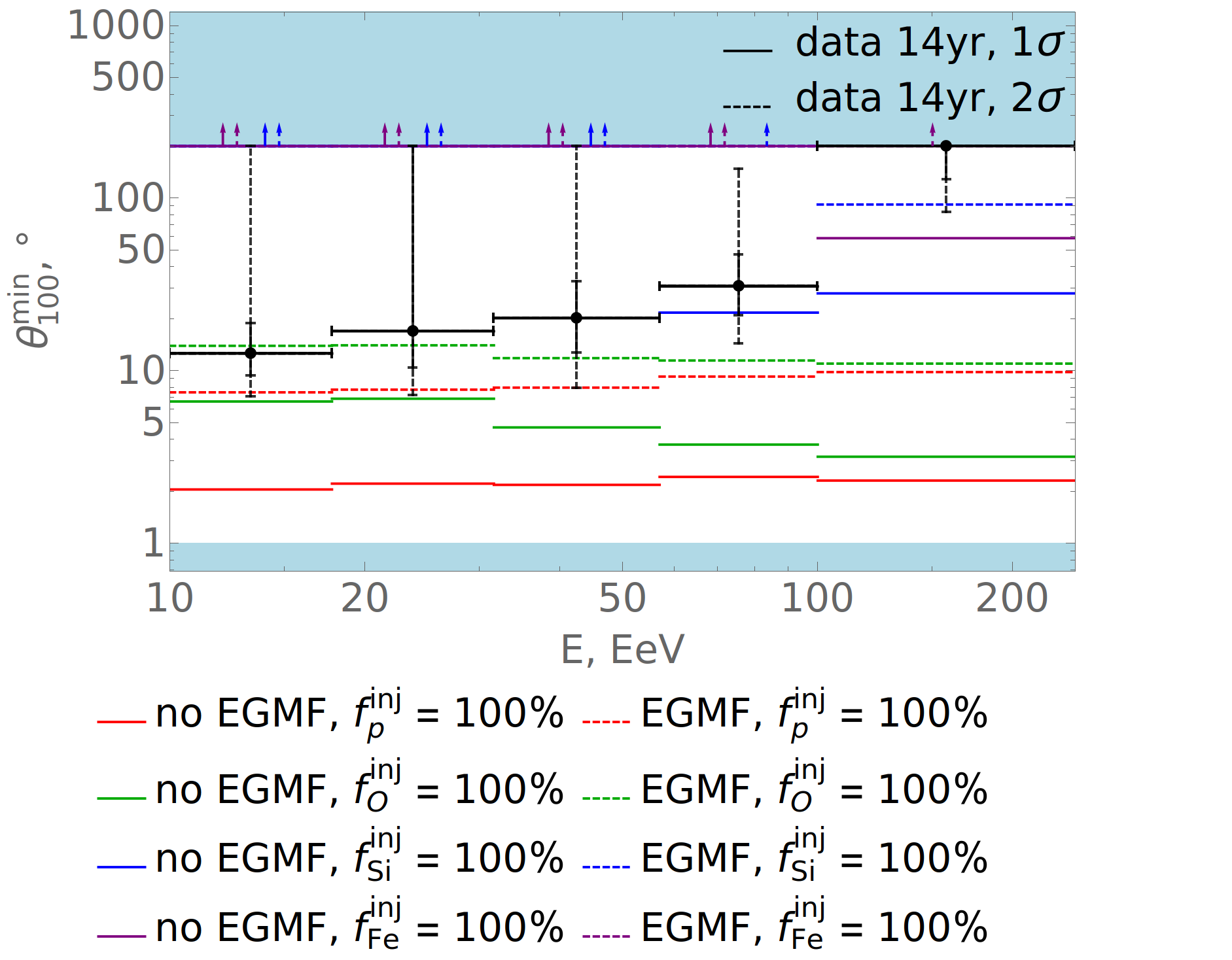}
\caption{
\label{fig:rails-MF}
The test statistics for the data compared to various pure nuclei injected
composition models.
{\it Left panel:} results for two different regular GMF models.
{\it Right panel:} results without EGMF and with extremely strong EGMF.}
\end{figure*}

The colored lines in Fig.~\ref{fig:rails} show predictions for different
composition models which should be compared to the data. With our assumptions
and zero EGMF the pure proton composition (red line) is not compatible with the data as
it predicts $\theta_{100}^{\rm min}\lesssim 2^\circ$ in all energy bins. The
injected light or intermediate composition is also incompatible with the data
as in this case the flux is dominated by secondary protons. At the same time
the data are compatible with the injected silicon at all energies except $E >
100$~EeV and with injected iron at all energies except $E \gtrsim 56$~EeV.
The Auger best-fit model is compatible with the data at $2\sigma$ level.
In general, one can see a trend: the preference for 
heavier composition at $10 < E \lesssim 18$~EeV changes in favor of a lighter one at $56
\lesssim E < 100$~EeV, while at $E > 100$~EeV the data prefer a very
heavy composition --- even beyond iron.

%Uncertainties
We turn now to the discussion of uncertainties affecting these results, of which the most important are those related to the magnetic fields, the experimental
energy scale and the injection spectrum.
In our setup all these uncertainties affect only the positions of model lines shown in Fig.~\ref{fig:rails}. The injection spectrum uncertainty was
tested by varying the spectrum parameters within $\pm 1 \sigma$ around their best fit values.
This variation was found to have negligible impact on the results, 
see Ref.~\cite{TelescopeArray:2024buq} for details.

To estimate the effect of GMF uncertainty we generate new mock sets, this time   
assuming the regular GMF model of Ref.~\cite{Jansson:2012pc}. Note that the
UHECR deflections in both models are similar in magnitude but substantially
differ in direction. The comparison of predicted values of $\theta_{100}^{\rm min}$
is shown in Fig.~\ref{fig:rails-MF}, left panel, for the same
composition models as in Fig.~\ref{fig:rails}. One can see that the predicted
values of $\theta_{100}^{\rm min}$ are quite close in almost all cases,
so that the change of the GMF model does not change the level
of compatibility of the composition models with the data. 

The EGMF is more uncertain than GMF. To estimate its impact on the results,
additional assumptions are required.  In general, there are three possible
regimes where EGMF may affect the UHECR deflections. First, there could be an
intergalactic magnetic field {\it IGMF} in voids of Large Scale Structure.  If
its origin is not cosmological its correlation length is expected not to
exceed $\sim 1$~Mpc~\cite{Durrer:2013pga}. Then its strength is bounded from
above as $B_{EGMF} < 1.7$~nG~\cite{Pshirkov:2015tua} and UHECR deflections are
described by a uniform smearing~\cite{Bhattacharjee:1998qc}. It is
straightforward to implement such a smearing into our simulation of mock sets. 
In the opposite case of the IGMF of cosmological origin, its amplitude is 
constrained to be $B \lesssim 0.05$~nG for any correlation
lengths~\cite{Jedamzik:2018itu}, that leads to deflections negligible
comparing to that in the GMF.
Finally, the IGMF can be negligible but there could be an EGMF in a local
extragalctic structures such as a local filament. There is no
observational bounds on such fields; however, constrained astrophysical
simulations predict its strength in the range $0.3 < B < 3$~nG in the $\sim 5$~Mpc
vicinity of our Galaxy~\cite{Hackstein:2017pex}.  Even in the conservative
case the expected deflections in such a field would be several times smaller
than the maximum possible deflections in IGMF.

Given all these considerations we test the possible effect of EGMF
conservatively assuming the highest allowed parameters for non-cosmological
field~\cite{Pshirkov:2015tua}: $B_{EGMF} = 1.7$~nG and $\lambda_{EGMF} = 1$~Mpc.
This may lead to deflections as high as $7^\circ$ for protons at 100 EeV. We
are simulating such deflections by an additional direction-independent smearing of sources
that scales according to the primary particle charge and energy.  The results
including both GMF and EGMF are shown in Fig.~\ref{fig:rails-MF}, right panel,
in comparison with the zero EGMF case. As one can see from the plot, the
inclusion of the maximum allowed EGMF significantly increases the value of 
$\theta_{100}^{\rm min}$ in all models and makes even the pure proton composition compatible
with the data in lower energy bins at the $2\sigma$ level.
In the last bin corresponding to $E>100$~EeV, this increase is
not sufficient except in the case of pure iron composition which becomes fully
compatible with the data.

The impact of the uncertainty related to the systematic uncertainty of the experiment's energy scale
is of the same order or smaller than the impact of the GMF uncertainty.
More detailed discussion of all the mentioned uncertainties is given in Ref.~\cite{TelescopeArray:2024buq}.

%Discussion
The interpretation of the results differs significantly depending on the assumed
deflections in EGMF, while the difference due to the GMF assumptions is
subdominant.  As it was mentioned, in the case of negligible EGMF the data prefer
a heavy composition at low energies, a relatively light one at $56 \lesssim E <
100$~EeV and a very heavy one (beyond iron) at $E > 100$~EeV. The latter result is in agreement with Ref.~\cite{TelescopeArray:2023sbd}, which finds that the TA highest energy event is not correlated with the LSS unless its deflection is very large. In the case of
extreme EGMF the data is consistent with both heavy and intermediate composition at $E < 100$~EeV.
In particular oxygen and even proton compositions became more compatible with the data at $E \lesssim 56$~EeV.

Importantly, the evidence of heavy composition at $E > 100$~EeV
survives the assumption of even extremely strong EGMF, while the light or
intermediate composition remains in tension with the data.
For instance, to reconcile the proton or helium composition with the data at $E > 100$~EeV at least at the $2\sigma$ level the EGMF should be stronger than 20~nG for $\lambda = 1$~Mpc, that is far beyond the upper limit discussed earlier.
It is also interesting that pure silicon is compatible with data from $10$~EeV
up to $100$~EeV irrespective of the EGMF.

In conclusion, an important comment concerning the interpretation of our results
in the low energy bins is in order. The logic here can be inverted:
taking at face value the light or intermediate composition measured at $10 \lesssim E \lesssim 50$~EeV
by the fluorescence experiments~\cite{TelescopeArray:2018xyi, PierreAuger:2014gko},
our results implying relatively large UHECR deflections at these energies point
toward the existence of a strong EGMF close to the current experimental limit.
The quantitative discussion of this observation will be given elsewhere.

\section*{Acknowledgements}
The authors would like to thank the former member of the Telescope Array collaboration Armando di Matteo, who kindly provided the simulations of UHECR propagation and respective fits of attenuation curves for the purposes of this study.

The Telescope Array experiment is supported by the Japan Society for
the Promotion of Science(JSPS) through
Grants-in-Aid
for Priority Area
%"Highest Energy Cosmic Rays"
431,
for Specially Promoted Research
%``Extreme Phenomena in the Universe Explored by Highest Energy Cosmic Rays''
%Grant Number
JP21000002,
%Grant-in-Aid
for Scientific  Research (S)
%"Quest for the unified picture of the explosion mechanism of supernovae and the central engine of gamma-ray bursts"
%Grant Number
JP19104006,
%Grant-in-Aid
for Specially Promoted Research
%"Extended Telescope Array Experiment - Nearby Extreme Universe Elucidated by Highest-energy Cosmic Rays"
%Grant Number
JP15H05693,
%Grant-in-Aid
for Scientific  Research (S)
%Grant Number
JP19H05607,
%Grant-in-Aid
for Scientific  Research (S)
%"Study of the ultra high energy cosmic ray source evolution by detailed measurement of cosmic rays in the wide energy range"
%Grant Number
JP15H05741,
%Grant-in-Aid
for Science Research (A)
%Grant Number
JP18H03705,
%Grant-in-Aid
for Young Scientists (A)
%"hoge hoge"
%Grant Number
JPH26707011,
and for Fostering Joint International Research (B)
%"Search for Ultra-High Energy Cosmic Ray origin using the extended Telescope Array experiment"
%Grant Number
JP19KK0074,
by the joint research program of the Institute for Cosmic Ray Research (ICRR), The University of Tokyo;
by the Pioneering Program of RIKEN for the Evolution of Matter in the Universe (r-EMU);
by the U.S. National Science
Foundation awards PHY-1806797, PHY-2012934, and PHY-2112904, PHY-2209583, PHY-2209584, and PHY-2310163, as well as AGS-1613260, AGS-1844306, and AGS-2112709;
by the National Research Foundation of Korea
% \linebreak
(2017K1A4A3015188, 2020R1A2C1008230, \& 2020R1A2C2102800) ;
%\linebreak
by the Ministry of Science and Higher Education of the Russian Federation under the contract 075-15-2024-541, IISN project No. 4.4501.18 by the Belgian Science Policy under IUAP VII/37 (ULB), by the European Union and Czech Ministry of Education, Youth and Sports through the FORTE project No. CZ.02.01.01/00/22\_008/0004632, and by the Simons Foundation (00001470, NG). This work was partially supported by the grants of The joint research program of the Institute for Space-Earth Environmental Research, Nagoya University and Inter-University Research Program of the Institute for Cosmic Ray Research of University of Tokyo. The foundations of Dr. Ezekiel R. and Edna Wattis Dumke, Willard L. Eccles, and George S. and Dolores Dor\'e Eccles all helped with generous donations. The State of Utah supported the project through its Economic Development Board, and the University of Utah through the Office of the Vice President for Research. The experimental site became available through the cooperation of the Utah School and Institutional Trust Lands Administration (SITLA), U.S. Bureau of Land Management (BLM), and the U.S. Air Force. We appreciate the assistance of the State of Utah and Fillmore offices of the BLM in crafting the Plan of Development for the site.  We thank Patrick A.~Shea who assisted the collaboration with valuable advice and supported the collaboration’s efforts. The people and the officials of Millard County, Utah have been a source of steadfast and warm support for our work which we greatly appreciate. We are indebted to the Millard County Road Department for their efforts to maintain and clear the roads which get us to our sites. We gratefully acknowledge the contribution from the technical staffs of our home institutions. An allocation of computing resources from the Center for High Performance Computing at the University of Utah as well as the Academia Sinica Grid Computing Center (ASGC) is gratefully acknowledged.

%\bibliographystyle{unsrturl}
\bibliography{ref.bib}

\end{document}